\documentclass[11pt]{article}

\usepackage[T1]{fontenc}
\usepackage[utf8]{inputenc}
\usepackage{lmodern}
\usepackage[margin=1in]{geometry}
\usepackage{amsmath,amssymb,amsfonts}
\usepackage{graphicx}
\usepackage{booktabs}
\usepackage{array}
\usepackage{tabularx}
\usepackage{multirow}
\usepackage{url}
\usepackage[hidelinks]{hyperref}
\usepackage{xcolor}
\usepackage{cite}
\usepackage{caption}
\usepackage{setspace}
\usepackage{enumitem}
\usepackage{float}
\usepackage{microtype}
\usepackage{authblk}

\graphicspath{{figures/}}
\hypersetup{
  pdftitle={Architectural Isolation as a Timing Safety Primitive for Edge AI Medical Devices: Controlled Experimental Evidence on a Shared-Silicon Platform},
  pdfauthor={Akul Mallayya Swami},
  pdfkeywords={edge AI, inference stability, timing safety, architectural isolation, STER, SaMD},
  pdfsubject={Research preprint}
}

\title{Architectural Isolation as a Timing Safety Primitive for Edge AI Medical Devices: Controlled Experimental Evidence on a Shared-Silicon Platform}

\author[1]{Akul Mallayya Swami}
\affil[1]{Varian Medical Systems, A Siemens Healthineers Company, Palo Alto, CA 94304 USA\\
Email: \texttt{swami.akul@alumni.uml.edu}\\
ORCID: 0009-0003-9549-5543}

\date{April 17, 2026}

\begin{document}
\maketitle

\begin{center}
\small
This work has been submitted to IEEE Embedded Systems Letters for possible publication.
Copyright may be transferred without notice,
after which this version may no longer be accessible.\\[0.4em]
This work was conducted independently and does not represent the views of
Varian Medical Systems or Siemens Healthineers.\\[0.4em]
\end{center}

\begin{abstract}
A system can satisfy accuracy-based validation, maintain output stability
(Safety-Threshold Exceedance Rate, STER, equal to zero), and still violate timing
constraints under deployment load. These are structurally independent properties that current
pre-market validation protocols often do not operationalize at the inference layer. This letter demonstrates their independence
through a controlled same-hardware experiment: identical MobileNetV2 models are
evaluated under identical adversarial load on two execution paths of the same NVIDIA
Jetson Orin Nano Super, a dedicated graphics processing unit (GPU) accelerator
(TensorRT FP16, half-precision floating point) and a general-purpose central processing
unit (CPU) (Open Neural Network Exchange (ONNX) Runtime FP32, single-precision
floating point). Both paths maintain STER$=0$; the CPU path (ONNX Runtime FP32) degrades 7.2$\times$
under combined load (mean latency 9.8$\times$ higher than the GPU path (TensorRT FP16),
which maintains latency below 11\,ms), breaching the 10\,Hz clinical cycle budget by 65\%. Joint STER
and latency verification is proposed as a candidate method for operationalizing
U.S. Food and Drug Administration (FDA) Draft Guidance FDA-2024-D-4488 robustness
requirements at the inference layer, subject to regulatory review and clinical validation.
\end{abstract}


\section{Introduction}

FDA pre-market validation of AI-enabled medical devices evaluates
model accuracy (sensitivity, specificity, area under the receiver operating curve (AUC))
under controlled, zero-contention conditions~\cite{fda2025}. This framework does not
characterize whether inference outputs remain within bounds, or whether inference
timing is preserved within clinical cycle budgets, under realistic deployment load.
A 2025 cross-sectional study of 950 FDA-cleared AI-enabled medical devices identified
60 devices subject to 182 recall events; diagnostic or measurement errors were the
leading cause, and 43\% of recalls occurred within the first year of
clearance~\cite{lee2025}.
International Electrotechnical Commission (IEC)~62304 Clause~6.3.2 requires software
timing verification under foreseeable conditions~\cite{iec62304}, yet accuracy-based
certification cannot detect timing violations under deployment load.

This letter demonstrates a challenge in operationalizing Software as a Medical
Device (SaMD) robustness requirements: while FDA-2024-D-4488 requires assessment
under foreseeable conditions of use~\cite{fda2025}, accuracy-based certification
protocols do not jointly verify output stability and timing reliability at the
inference layer under deployment load. The safety criterion
is not latency magnitude but whether latency satisfies the clinical deadline; a change
in accuracy ($\Delta\text{acc}$) of less than 1.0\% does not indicate deadline
compliance. The experiment holds all hardware variables constant so that divergence arises from
the execution stack, the combined effect of architectural path, runtime, and
numerical precision, rather than from chip-to-chip or thermal variability. Results show STER$=0$ on both paths while CPU
$P_{99}$ (99th-percentile latency) exceeds the 10\,Hz budget by 65\% under combined
load. A joint STER and latency protocol, consistent with FDA-2024-D-4488, is described.

\section{Related Work}

Prior edge inference work addresses throughput maximization~\cite{hao2023},
GPU partitioning for temporal isolation on Jetson Orin (MPS, MIG, Green
Contexts)~\cite{marin2026}, and cross-device accuracy
benchmarking~\cite{baller2021,tobiasz2023}. These measure latency or accuracy in
isolation, not their joint characterization under shared-silicon contention;
\cite{marin2026} is closest in platform and concern but addresses timing
isolation between parallel GPU processes, not output stability as an independent
safety property, and does not include a CPU-path comparator.
Safety-oriented work on neural networks in medical devices~\cite{becker2019} addresses
model-level mechanisms but does not empirically characterize resource-contention-induced
timing failure as a distinct property from output correctness.
Established methods such as worst-case execution time (WCET) analysis~\cite{wilhelm2008}
address timing at the scheduler level but do not characterize inference output stability
under deployment-induced contention.
No prior work has used a same-hardware controlled design to empirically demonstrate
that output stability and timing reliability are orthogonal properties, or proposed
their joint measurement as a pre-market acceptance condition for SaMD.

\section{System Model and Formal Definitions}

\subsection{Output Stability Under Identical Inputs}

Let $f$ denote a deployed inference model and $\mathbf{x}_i$ a fixed test input.
Under stressor condition $c$, the per-inference output deviation is:
\begin{equation}
  \delta_i = \bigl\|\sigma(\mathbf{y}_i^c) - \bar{\sigma}_i\bigr\|_{\infty}
           = \max_k \bigl|\sigma(\mathbf{y}_i^c)_k - \bar{\sigma}_{i,k}\bigr|
  \label{eq:delta}
\end{equation}
where $\sigma(\mathbf{y}_i^c)$ is the softmax output under condition $c$ and
$\bar{\sigma}_i$ is the per-image reference at zero load.

\subsection{STER Definition}

\textbf{Definition~1} ($T^*$): $T^* = 0.05$, chosen to detect subthreshold
distribution shifts below typical clinical tolerance bands. Clinical validation
of this threshold for specific device classes is required~\cite{iso14971}.

\textbf{Definition~2} (STER): For $N$ inference activations:
\begin{equation}
  \text{STER} = \frac{1}{N}\sum_{i=1}^{N} \mathbf{1}[\delta_i > T^*]
  \label{eq:ster}
\end{equation}

Accuracy testing evaluates $\text{argmax}(\sigma(\mathbf{y}))$ and cannot detect
subthreshold distribution drift; a model shifting confidence by 4.99\% passes accuracy
testing. STER detects this class of silent deviations. STER is a candidate pre-market
metric, not an existing FDA-specified standard.

\subsection{Robustness Confirmation Condition}

The joint pre-market acceptance condition is:
\begin{equation}
  \text{STER}_c \leq \text{STER}_{\max}
  \quad\text{AND}\quad
  \text{latency}_{P_{99}} \leq T_n
  \label{eq:joint}
\end{equation}
where $\text{STER}_{\max}=0$, $\text{latency}_{P_{99}}$ is the 99th-percentile
inference latency, and $T_n=100$\,ms is the nominal cycle period at 10\,Hz (a
representative constraint; appropriate rates are device-specific). STER and latency
are orthogonal safety dimensions; both must be verified.

\subsection{Architectural Model}

\begin{figure}[t]
  \centering
  \includegraphics[width=0.9\textwidth]{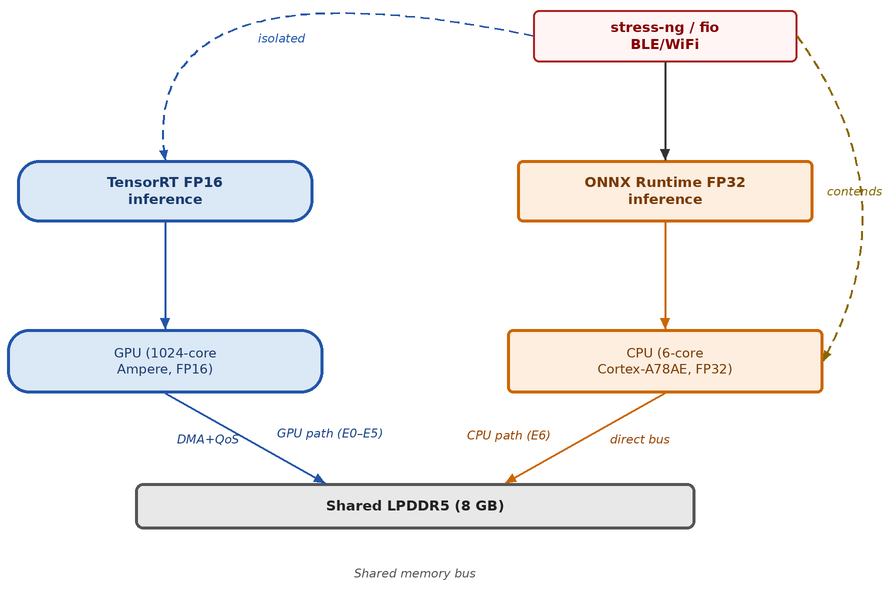}
  \caption{Jetson Orin Nano Super execution paths. GPU path (TensorRT FP16) uses
dedicated DMA channels with QoS bandwidth reservation. CPU path (ONNX Runtime FP32)
competes directly for bus bandwidth with stressor workloads.}
  \label{fig:arch}
\end{figure}

The Jetson Orin Nano Super uses shared Low-Power Double Data Rate 5 (LPDDR5) memory;
CPU, GPU, and memory controller compete for the same bus (Fig.~\ref{fig:arch}). The
GPU path (TensorRT FP16) offloads inference to the 1024-core Ampere GPU via dedicated
Direct Memory Access (DMA) paths with Quality of Service (QoS) bandwidth reservation.
The CPU path (ONNX Runtime FP32) runs on the 6-core Arm Cortex-A78AE, competing
directly with stressor workloads for CPU time, cache, and dynamic random-access memory
(DRAM) bandwidth.

\section{Hardware Platform and Experimental Protocol}

\subsection{Hardware Platform}

Both execution paths run MobileNetV2~\cite{sandler2018} ($224{\times}224$
red-green-blue (RGB), ImageNet 1000-class head) on the same 500-image fixed test set.
The GPU path runs TensorRT FP16~\cite{tensorrt2024} at a nominal 10\,Hz
($T_n=100$\,ms); the CPU path runs MobileNetV2 FP32 via ONNX
Runtime~\cite{onnxruntime2024} on the same hardware. Same-device testing eliminates
chip-to-chip variation and thermal confounds~\cite{marin2026}. No thermal throttling
events were recorded (\texttt{tegrastats} monitoring throughout).

\begin{table}[t]
\caption{Hardware Platform Summary}
\label{tab:hardware}
\centering
\small
\begin{tabularx}{\textwidth}{@{}>{\raggedright\arraybackslash}p{0.42\textwidth} >{\raggedright\arraybackslash}X >{\raggedright\arraybackslash}p{0.22\textwidth}@{}}
\toprule
\textbf{Component} & \textbf{Role} & \textbf{Key Spec.} \\
\midrule
Jetson Orin Nano Super (GPU) & Inference host & TensorRT FP16 \\
Jetson Orin Nano Super (CPU) & Same hardware comparator path & ONNX Runtime FP32 \\
nRF52840 DK + TI SensorTag & BLE stressor & BLE 5.0, 6 connections \\
GL-AX1800 WiFi 6 Router & RF environment & 2.4\,GHz \\
\bottomrule
\end{tabularx}
\end{table}

\subsection{Experimental Protocol}

Each trial: $N=500$ inference activations; softmax $\sigma(\mathbf{y}_i)$ and latency
(via \texttt{time.perf\_counter()}) recorded per activation; $\delta_i$
computed per~(\ref{eq:delta}); STER computed per~(\ref{eq:ster}). All CPU inferences
used \texttt{CPUExecutionProvider} in deterministic mode. Trial counts ($N=500$;
10 trials per stressor level) provide stable mean and $P_{99}$ estimates.
The 9.8$\times$ mean latency difference (10.6\,ms GPU vs. 104.0\,ms CPU) is large
relative to within-condition variability (CPU SD $= 5.7$\,ms under combined load),
indicating a robust effect; confidence intervals and formal hypothesis tests are
deferred to a full-length follow-up study.

\begin{table}[t]
\caption{Experiment Summary}
\label{tab:experiments}
\centering
\small
\begin{tabularx}{\textwidth}{@{}>{\raggedright\arraybackslash}p{0.10\textwidth} >{\raggedright\arraybackslash}p{0.16\textwidth} X >{\centering\arraybackslash}p{0.08\textwidth}@{}}
\toprule
\textbf{ID} & \textbf{Stressor} & \textbf{Protocol} & \textbf{IEC}$^a$ \\
\midrule
E0 & Baseline & Both paths, zero load. Reference $\bar{\sigma}_i$ captured. 10$\times$500 inferences. & B+C \\
E1 & CPU stress & \texttt{stress-ng}~\cite{king2023} at 25/50/75/100\%. Both paths. 10$\times$500 inferences per level. & B+C \\
E2--E4 & Isolation & Memory (25--90\%), GPU co-tenancy (0--4), network I/O (BLE 0--6 + WiFi). GPU only; STER$=0$. & B+C \\
E5 & Combined & CPU 75\%, memory 50\%, BLE 4 + fio. Both paths. 10$\times$500 inferences. & B+C \\
E6 & CPU char. & CPU path only. E0 + E1 + E5 conditions. 10$\times$500 inferences. & B/C \\
\bottomrule
\multicolumn{4}{@{}l@{}}{\scriptsize $^a$ IEC~62304: B = Clause 5.7 (integration); C = Clause 5.8 (system testing).}
\end{tabularx}
\end{table}

\section{Results}
\label{sec:results}

\subsection{E0: Baseline Calibration}

Both paths establish zero-load baselines (Table~\ref{tab:baseline}). GPU latency
(10.6\,ms mean; 10.9\,ms $P_{99}$) reflects end-to-end pipeline time; CPU latency
(14.5\,ms mean) reflects pure ONNX Runtime execution. GPU floating-point variance
produces non-zero $\delta_{\text{mean}}$ but zero STER; CPU is arithmetically
deterministic. Top-1 predictions are recorded as a control variable only; the test
set uses Tiny ImageNet images against the full ImageNet-1000 head, so absolute
accuracy is not a dependent measure. Equality of top-1 distributions across
conditions ($\Delta\text{acc}<1.0\%$) confirms that accuracy-based validation would
not flag any tested condition.

\begin{table}[t]
\caption{E0 Baseline. $\delta_{\text{mean}}$: mean output deviation from \eqref{eq:delta}; Lat. SD: per-trial standard deviation; $P_{99}$: 99th-percentile latency.}
\label{tab:baseline}
\centering
\small
\begin{tabularx}{\textwidth}{@{}>{\raggedright\arraybackslash}p{0.28\textwidth} >{\centering\arraybackslash}p{0.10\textwidth} >{\centering\arraybackslash}p{0.14\textwidth} >{\centering\arraybackslash}p{0.13\textwidth} >{\centering\arraybackslash}p{0.11\textwidth} >{\centering\arraybackslash}p{0.11\textwidth}@{}}
\toprule
\textbf{Path} & \textbf{STER} & $\boldsymbol{\delta_{\text{mean}}}$ & \textbf{Lat. mean} & \textbf{Lat. SD} & \textbf{$P_{99}$} \\
\midrule
GPU (TRT FP16)$^\dagger$ & 0.0000 & 0.0182 & 10.6\,ms & 0.2\,ms & 10.9\,ms \\
CPU (ONNX FP32) & 0.0000 & 0.0000 & 14.5\,ms & 0.2\,ms & 18.6\,ms \\
\bottomrule
\multicolumn{6}{@{}p{\textwidth}@{}}{\scriptsize $^\dagger$ End-to-end (preprocess + infer + softmax), $N = 5000$.}
\end{tabularx}
\end{table}

\subsection{E1: CPU Stress and Timing Divergence}

Both paths maintain STER$=0.0000$ at all CPU load levels. The GPU path is invariant
($\approx 10.6$\,ms); the CPU path reaches 73.7\,ms at 75\% load (5.1$\times$
baseline). TensorRT FP16 executes on the GPU, isolated from saturated CPU cores;
ONNX Runtime FP32 competes for CPU time slices and cache bandwidth.

\begin{table}[t]
\caption{E1 CPU Stress: Both Paths (10 Trials $\times$ 500 Inferences Per Level)}
\label{tab:e1}
\centering
\small
\begin{tabularx}{\textwidth}{@{}>{\centering\arraybackslash}p{0.16\textwidth} >{\centering\arraybackslash}p{0.16\textwidth} >{\centering\arraybackslash}p{0.16\textwidth} >{\centering\arraybackslash}p{0.16\textwidth} >{\centering\arraybackslash}p{0.16\textwidth}@{}}
\toprule
\textbf{CPU Load} & \textbf{GPU STER} & \textbf{GPU Lat.} & \textbf{CPU STER} & \textbf{CPU Lat.} \\
\midrule
25\% & 0.0000 & $\approx 10.6$\,ms & 0.0000 & 33.4\,ms \\
50\% & 0.0000 & $\approx 10.6$\,ms & 0.0000 & 57.8\,ms \\
75\% & 0.0000 & $\approx 10.6$\,ms & 0.0000 & 73.7\,ms \\
100\% & 0.0000 & $\approx 10.6$\,ms & 0.0000 & 70.9\,ms \\
\bottomrule
\end{tabularx}
\end{table}

\subsection{E2--E4: Isolation Stressors (GPU Path Only)}

Memory pressure (25--90\% LPDDR5 fill), GPU co-tenancy (0--4 concurrent TensorRT
workers), and network I/O (BLE 0--6 + WiFi) all produce STER$=0.0000$ with
$\delta_{\text{mean}}$ indistinguishable from E0. GPU DMA isolation and QoS
bandwidth reservation prevent all stressor classes from perturbing outputs or latency
($+88\%$ interrupt rate at BLE$=4$ had no effect).

\subsection{E5: Combined Realistic Load}

\begin{figure}[t]
  \centering
  \includegraphics[width=0.82\textwidth]{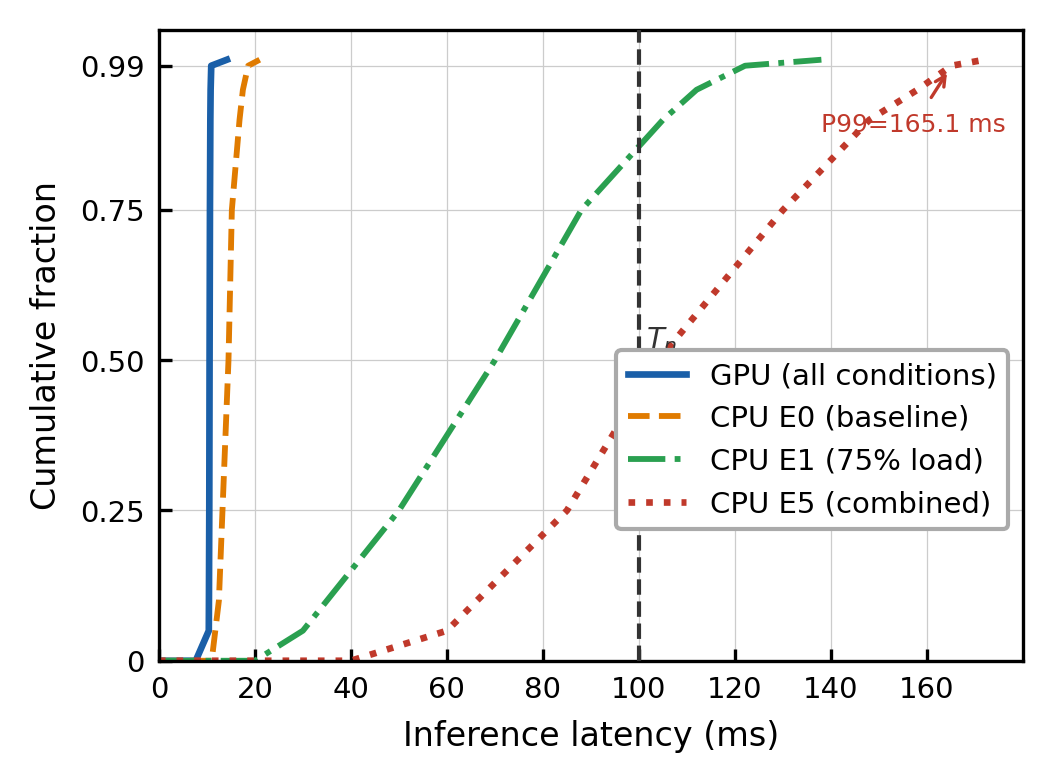}
  \caption{Empirical cumulative distribution function (CDF) of inference latency for
GPU and CPU paths under baseline (E0), CPU stress (E1 75\%), and combined load (E5).
GPU $P_{99}=10.9$\,ms across all conditions. CPU path exceeds $T_n=100$\,ms at the
75th percentile under E5 ($P_{99}=165.1$\,ms); undetected by accuracy-based
validation ($\Delta\text{acc}<1.0\%$).}
  \label{fig:cdf}
\end{figure}

Under simultaneous CPU 75\%, memory 50\% fill, 4 BLE connections + fio disk write,
the GPU path maintains STER$=0.0000$ and latency below 11\,ms (mean 10.6\,ms),
9.8$\times$ lower than the CPU path. The CPU path maintains STER$=0.0000$ but
reaches 104.0\,ms mean, 165.1\,ms $P_{99}$, exceeding the 10\,Hz budget by 65\%
(Fig.~\ref{fig:cdf}).

\subsection{E6: CPU Contention Characterization}

STER$=0.0000$ and $\delta_{\text{mean}}=0.0000$ across all 30,000 inferences;
\texttt{CPUExecutionProvider} uses deterministic ARM NEON Single Instruction Multiple
Data (SIMD) operations. $P_{99}=165.1$\,ms under combined load means 1\% of
inferences exceed the clinical budget by 65\%. The GPU path maintains below 11\,ms
under identical load (9.8$\times$ lower), consistent with timing isolation as a
property of GPU-accelerated execution stacks under the tested conditions.

\begin{table}[t]
\caption{E6 CPU-Only Contention: ONNX Runtime FP32 (10 Trials $\times$ 500 Inferences)}
\label{tab:e6}
\centering
\small
\begin{tabularx}{\textwidth}{@{}>{\raggedright\arraybackslash}p{0.18\textwidth} >{\centering\arraybackslash}p{0.10\textwidth} >{\centering\arraybackslash}p{0.13\textwidth} >{\centering\arraybackslash}p{0.14\textwidth} >{\centering\arraybackslash}p{0.12\textwidth} >{\centering\arraybackslash}p{0.11\textwidth}@{}}
\toprule
\textbf{Condition} & \textbf{STER} & $\boldsymbol{\delta_{\text{mean}}}$ & \textbf{Lat. mean} & \textbf{Lat. SD} & \textbf{$P_{99}$} \\
\midrule
Zero load & 0.0000 & 0.0000 & 14.5\,ms & 0.2\,ms & 18.6\,ms \\
CPU 25\% & 0.0000 & 0.0000 & 33.4\,ms & 1.8\,ms & 111.4\,ms \\
CPU 50\% & 0.0000 & 0.0000 & 57.8\,ms & 3.2\,ms & 115.8\,ms \\
CPU 75\% & 0.0000 & 0.0000 & 73.7\,ms & 4.1\,ms & 122.1\,ms \\
CPU 100\% & 0.0000 & 0.0000 & 70.9\,ms & 3.8\,ms & 117.0\,ms \\
\textbf{Combined}$^\dagger$ & \textbf{0.0000} & \textbf{0.0000} & \textbf{104.0\,ms} & \textbf{5.7\,ms} & \textbf{165.1\,ms} \\
\bottomrule
\multicolumn{6}{@{}p{\textwidth}@{}}{\scriptsize $^\dagger$ Exceeds $T_n = 100$\,ms budget by 65\%.}
\end{tabularx}
\end{table}

\section{Discussion}

The finding is not that timing degrades under load, but that existing accuracy-based
validation cannot detect this degradation, as demonstrated empirically here.

\subsection{Controlled Design as Methodological Contribution}

The same-hardware design controls for all hardware-level confounds; CPU-path timing
degradation of 7.2$\times$ (14.5 to 104\,ms mean) and the 9.8$\times$ GPU/CPU
contrast under combined load reflect the combined effect of execution path architecture,
runtime (TensorRT vs. ONNX), and precision (FP16 vs. FP32).

\subsection{Timing as the Dominant Failure Mode}

STER$=0.0000$ on both paths is not a null result: it is the necessary condition
that makes the timing finding interpretable. Without confirmed output stability,
an observed CPU-path latency increase could reflect output-level degradation (slower
inference due to numerical instability) rather than pure scheduling pressure.
STER$=0$ rules this out: scheduling pressure delays computation without perturbing
arithmetic, making timing degradation the sole identifiable safety-relevant divergence
(Fig.~\ref{fig:scatter}).

\begin{figure}[t]
  \centering
  \includegraphics[width=0.82\textwidth]{fig3_ster_vs_latency.png}
  \caption{STER vs. mean latency for all experimental conditions. Both GPU and CPU
paths maintain STER$=0$ (below $T^*=0.05$) regardless of latency. GPU clusters at
10.6\,ms across all conditions; CPU spans 14.5--104.0\,ms, breaching $T_n=100$\,ms
while STER remains zero.}
  \label{fig:scatter}
\end{figure}

\subsection{Operationalizing Pre-Market Robustness Requirements}

A manufacturer using accuracy-only testing would clear both paths as equivalent;
only joint condition~(\ref{eq:joint}) distinguishes them: $\Delta\text{acc}<1.0\%$
on both paths while the CPU path breaches the 10\,Hz budget by 65\%.
This limitation is structural: accuracy evaluates the argmax of a single
inference output and carries no information about the temporal spacing of successive
outputs, yet FDA-2024-D-4488 requires robustness assessment under foreseeable conditions of use.
A device operating at 10\,Hz produces one inference activation every 100\,ms;
if that activation requires 165\,ms, the clinical loop stalls regardless of whether
the output is numerically correct.
FDA-2024-D-4488 requires robustness assessment under foreseeable conditions of
use~\cite{fda2025}; the joint STER + latency condition operationalizes this
requirement at the inference layer. Neither STER alone nor latency alone is
sufficient: STER$=0$ on the CPU path confirms output correctness while the
latency violation confirms deployment infeasibility, and only their conjunction
exposes the failure.

\subsection{STER as Output Stability Certification}

STER certifies that the full output distribution $\sigma(\mathbf{y})$ remains within
$T^*$ under all tested conditions. Accuracy (top-1 class) cannot provide this
certification as it does not observe subthreshold distribution shifts. STER$=0$ on
both paths isolates timing degradation as the sole safety-relevant divergence.
This isolation is the methodological contribution: without a confirmed output-stability
metric, a reviewer cannot distinguish between two failure hypotheses: (1) the model
output degrades under contention, or (2) only timing degrades. STER$=0$ rules out
hypothesis (1) across all tested conditions, making the timing failure the clean
and attributable result.
The $T^*=0.05$ threshold is a proposed value requiring clinical validation for
specific device classes; tighter thresholds would not change the finding since
$\delta_{\text{max}}$ never exceeded 0.025 across any of the 107,500 inference
activations. Appropriate thresholds should be established through risk analysis
per ISO~14971~\cite{iso14971}.
Future work should establish clinically validated thresholds for specific device
classes, and extend the protocol to non-deterministic runtimes where $\delta>0$
may appear under contention even without argmax changes.

\subsection{Limitations}

Results are specific to MobileNetV2 at 10\,Hz on Jetson Orin Nano Super under
JetPack~6. $T^*=0.05$ is a proposed threshold requiring clinical validation for
specific device classes. The GPU and CPU paths differ in runtime (TensorRT vs. ONNX),
precision (FP16 vs. FP32), and measurement scope (end-to-end vs. runtime-only);
the 9.8$\times$ latency difference reflects the combined effect of these factors and
cannot be attributed solely to architectural isolation. The STER$=0$ finding is
precision-independent and holds across both execution stacks. This study does not
evaluate preprocessing latency, real-time operating system (OS) scheduling,
multi-model pipelines, or a complete regulated clinical workflow.

\section{Conclusion}

Output correctness and timing reliability are independent safety properties of edge
AI inference pipelines, as demonstrated by comparing two complete execution stacks
(TensorRT GPU FP16 vs. ONNX CPU FP32) on the same hardware platform. Both paths maintain STER$=0$ across all conditions; the GPU
path maintains latency below 11\,ms while the CPU path degrades 7.2$\times$ to
104\,ms mean ($P_{99}=165$\,ms, 9.8$\times$ higher), breaching the 10\,Hz budget by
65\%. Accuracy-based pre-market testing would pass both paths without detecting this
timing failure. Joint STER + latency verification is proposed as a candidate method
for operationalizing FDA-2024-D-4488 robustness requirements at the inference layer,
consistent with IEC~62304~\cite{iec62304} and ISO~14971~\cite{iso14971}, and subject
to regulatory review and clinical validation for specific device classes.

\section*{Data Availability}
Experimental scripts, raw results, and analysis code are publicly available at
\url{https://github.com/akulswami/p2-edge-ai-samd}.

\end{document}